# Dynamics of Fe atoms in Fe-gluconate as seen by Mössbauer spectroscopy

S.M. Dubiel[*] and J. Cieślak

*AGH University of Science and Technology, Faculty of Physics and Applied Computer Science, al. Mickiewicza 30, 30-059 Krakow, Poland*

## Abstract

Fe-gluconate was studied by means of the Mössbauer spectroscopy in the temperature interval of 80-305K. The measured spectra were analyzed in terms of two subspectra and, alternatively, two distributions of the quadrupole splitting. The major component (~85%) was identified as due to high-spin $Fe^{2+}$ ions while the identification of the minor component (~15%) was not unique: some characteristics are in favour of the high-spin $Fe^{3+}$ state while other ones are consistent with the low-spin $Fe^{2+}$ state. Values of the Debye temperature were determined for both phases from a temperature dependence of the center shift as well as from that of the spectral area. The force constant for the high-spin ferrous atoms was evaluated to be equal to 44 N/m (243 cm$^{-1}$).

[*]Corresponding author: Stanislaw. Dubiel@fis.agh.edu.pl

## 1. Introduction

Ferrous gluconate is a metalorganic compound (a salt of the gluconic acid) containing iron in its ferrous (Fe(II) or $Fe^{2+}$) form. It has been used in the pharmacological industry for a production of medicaments, e. g. Apo® ferrous, Apotex®, Ascofer®, Fergon®, Ferralet®, Simron®, to name just a few, applied in treatments of iron deficiency anemias. Interestingly, it has also found applications in the metallurgical industry where it was revealed as an effective inhibitor for a carbon steel [1], and gluconate-based electrolytes were also successfully used to electroplate various metals [2] or alloys [3] . Concerning its medical applications, it is well known that an iron intake by humans depends on a number of factors, one of them being its redox state (ferrous or ferric) e. g. [4]. The bioavailability of the ferrous iron lies between 10 and 15% and is 3-4 times higher than the one of the ferric iron e. g. [5]. Consequently, a purity of the iron gluconate is of importance as far as its applications as iron supplementation is concerned. The ferrous iron is known to be prone to oxidation. In fact, Mössbauer effect measurements which are able to distinguish between different valence states of iron gave evidence that the ferrous iron constituted less than ~90% of the total iron present in the iron gluconate [6-10], the rest being the ferric iron. However, it must be remembered that the detectability of the two form of iron may be different. In other words the $Fe^{2+}/Fe^{3+}$ ratio obtained from the spectral area of the corresponding sub spectra does not necessarily reflect the real contribution of the two iron phases. In general, as discussed in detail elsewhere [11] for various $Fe^{2+}$- and $Fe^{3+}$-containing oxides, oxyhydroxides, silicates and carbonates, the so-called *f*-factor (recoil free fraction) which is responsible for the detectability of a given phase is higher for the ferric iron assuming the both types of ions are embedded in the same matrix. The *f*-related dynamical quantity viz. the Debye temperature, $\theta_D$, can differ by more than 200K for the ferrous and the ferric ions [11]. In other words, the relative amount of the ferric phase derived from the spectral area may be overestimated. For the inorganic compounds mentioned above the overestimation may be as high as 15% for structurally related compounds [11]. This uncertainty is as high as the amount of the ferric phase in the Fe-gluconate as determined by the Mössbauer spectroscopy [6-10]. The issue of the minor ferric phase is further complicated by the fact that the structure of the Fe-gluconate is not precisely known. Recent studies performed on a high-purity sample by means of five different techniques did not give the definite answer in that matter [12]. In the Mösbauer spectroscopy, the distinction between the ferrous and the ferric spices is based on values of two spectral parameters: isomer shift (IS) and quadrupole splitting (QS). Their values depend not only on the valence state but also on the spin state i.e. low spin (LS) or high spin (HS). As reported elsewhere [13], one cannot uniquely identify the ferric ions as their *IS* and *QS* values overlap with those of $Fe^{2+}$(LS). In these circumstances, as remarked elsewhere [10], the minor iron phase seen in the Mössbauer spectra and commonly regarded as the ferric one may be , in fact, a ferrous low-spin phase. To shed more light on the issue, the present study aimed at determination of the Debye temperature, $\theta_D$, of both forms of iron was undertaken. It is known that the lattice dynamics depends on the crystal symmetry as well as on charges of vibrating atoms. As mentioned above the $\theta_D$-values are significantly higher for the ferric phase which follows from a higher charge of the former, hence a stronger coupling to the lattice. Therefore it seems of interest to determine the $\theta_D$-values for the two iron phases present in the Fe-gluconate. To our best knowledge no such data are available in the literature.

**2. Experimental**

**2.1. Sample, spectra measurements and analysis**

As a sample was used Fe-gluconate courtesy of the Chemistry and Pharmacy Cooperative ESPEFA (Krakow, Poland) which uses it for a production of an iron supplement Ascofer®. $^{57}$Fe-site Mössbauer spectra were recorded in a transmission geometry and in the temperature range of 80-305K using a standard spectrometer and a sinusoidal drive. The 14.4 keV gamma rays were supplied by a ~20 mCi Co/Rh source. The sample consisting of 200 mg Fe-gluconate in form of powder was distributed homogeneously on a surface of ~3 cm$^2$ i.e. the density of Fe was ~8 mg/cm$^2$. The spectra, examples of which can be seen in Fig. 1, had the same shape at all temperatures measured. They were analyzed with two procedures: (A) superposition of two doublets with Lorentzian shape of lines: one to account for the sub spectrum associated with the ferrous iron (higher intensity and splitting), and the other one (smaller intensity and splitting) to account for the ferric-HS or ferrous-LS phase, and (B) distribution of the quadrupole splitting, *QSD*. In the latter two distributions of *QS* with different isomer shifts were assumed. The spectra could have been successfully fitted with both procedures and the output of the B-procedure is shown in the right panel of Fig. 2. Here, the peak centered at *QS*≈3 mm/s corresponds to the ferrous (Fe$^{2+}$-HS) iron, and the one situated at *QS*≈0.8 mm/s represents the ferric -HS or the ferrous-LS iron. It can be seen that the former has a shoulder on its left side a feature that indicates a presence of the Fe$^{2+}$ ions experiencing a slightly weaker crystal field. The existence of such ions was also reported previously [6-8,10]. It should be also noticed that the width of the peak representing the minor iron phase is larger than the one associated with the major phase. This may be due to a less ordered structure of the former.

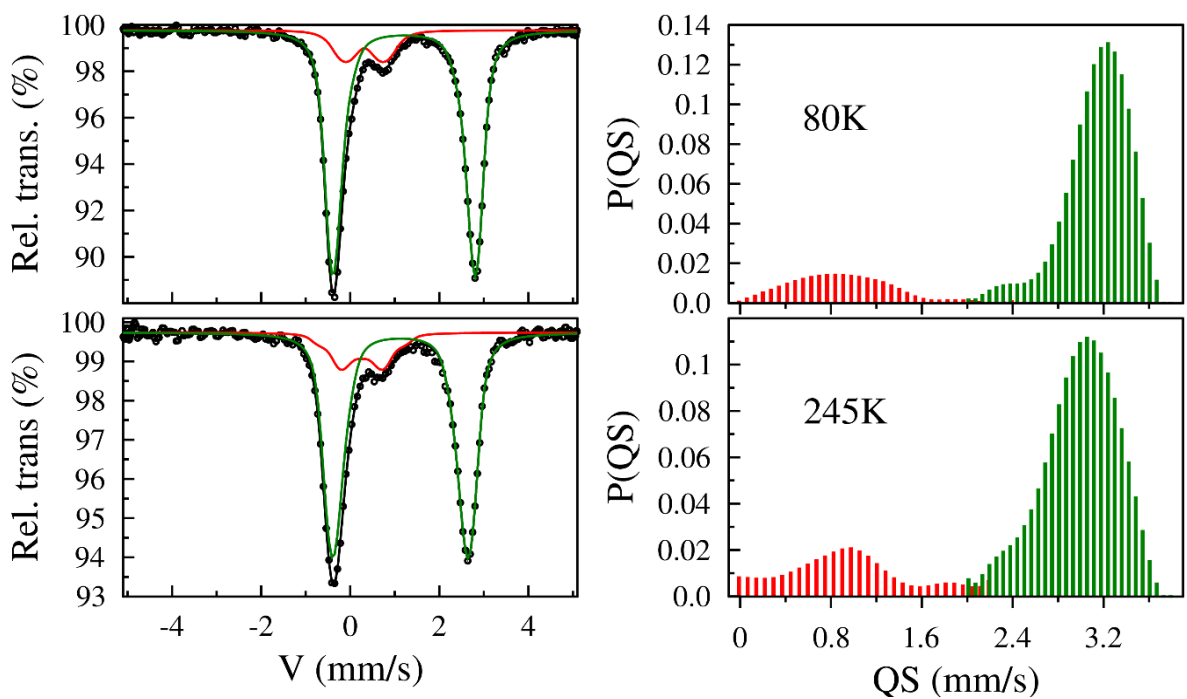

Fig. 1

$^{57}$Fe-site Mössbauer spectra (left panel) recorded at two different temperatures shown, and the corresponding distributions of the quadrupole splitting, *QS* (right panel). Small peak centered at *QS*≈0.8 mm/s represents the ferric-HS or the ferrous-LS iron, while the one centered at *QS*≈3 mm/s stands for the ferrous-HS phase.

Both procedures yielded spectral parameters pertinent to determination of the Debye temperature viz. a center shift, *CS*, and the spectral area, *A*, that is related to the recoil free fraction, *f*.

## 3. Results and discussion

### 3.1. Temperature dependence of center shift

The temperature dependence of *CS* can be expressed as follows:

$$CS(T) = IS(T) + SOD(T) \qquad (1)$$

Where *IS* is the isomer shift and *SOD* is the so-called second order Doppler shift. Assuming that in the first-order approximation the phonon spectrum can be described by the Debye model, and taking into account that the temperature dependence of *IS* is weak, so it can be neglected [14,15], the temperature dependence of *CS* can be related to the Debye temperature via the second term in eq. (1) as follows:

$$CS(T) = -\frac{3k_B T}{2mc}\left(\frac{3\Theta_D}{8T} + 3\left(\frac{T}{\Theta_D}\right)^3 \int_0^{\Theta_D/T} \frac{x^3}{e^x - 1}dx\right) \qquad (2)$$

Where *m* is the mass of a Fe atom, $k_B$ is the Boltzmann constant, *c* is the speed of light, and $x = \hbar\omega/kT$ ($\omega$ being frequency of vibrations).

The temperature dependence of the average center shift, *<CS>*, as derived from the two fitting procedures for the major and the minor sub spectra is presented in Fig. 2.

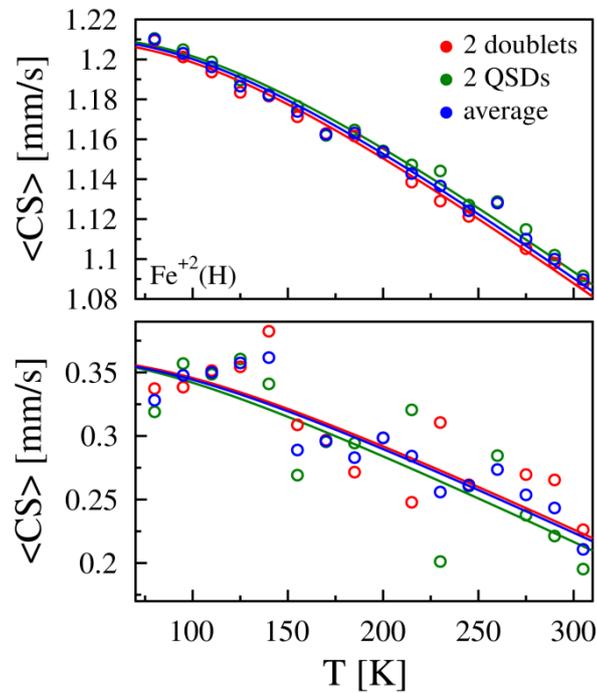

Fig. 2

Temperature dependence of the average center shift, <CS>, as obtained from the fitting procedures A and B as well as that of the arithmetic average over A and B (open symbols). The solid lines stand for the best fit in terms of eq. (2). The top panel is for the major ($Fe^{2+}$(H)) and the bottom one for the minor phase.

It can be seen that the data obtained from the major sub spectrum hardly depend on the fitting procedure while the ones derived from the minor spectrum exhibit a high degree of dispersion. To account for the differences in the <CS> values due to the fitting procedures, the arithmetic average over A and B (indicated in Table 1 as C) was calculated, too. The best fit of eq. (2) to the three sets of data yielded values of $\theta_D$ that are displayed in Table 1.

Table 1 The values of the Debye temperature as determined from the temperature dependence of (a) <CS>, $\theta_D$(CS), and (b) $ln(f/f_o)$, $\theta_D$(f) for the major and the minor iron phases.

|  | Major phase (HS-$Fe^{2+}$) | | | Minor phase | | |
|---|---|---|---|---|---|---|
|  | A | B | C | A | B | C |
| $\theta_D$ (CS) [K] | 438(23) | 460(21) | 445(18) | 367(171) | 315(166) | 357(111) |
| $\theta_D$ (f) [K] | 210(10) | 202(13) | 207(13) | 214(58) | 238(50) | 225(48) |

The large errors of the $\theta_D$(CS)-values found for the minor phase do not permit to conclude whether or not these values are different than those estimated for the major phase. In any case they are not significantly higher as expected for the ferric iron [11]. For Ferrum Lek, a commercially available iron-polymaltose complex, in which Fe-bearing ions are ferric, the value of $\theta_D$(CS)=502(24) K was found [12] comparing well with the values expected for the ferric iron [11].

## 3.2. Temperature dependence of spectral area

Alternatively, the Debye temperature can be determined from a temperature dependence of the *f*-factor. In the frame of the Debye model the *f*-$\theta_D$ relationship reads as follows:

$$f = \exp\left[\frac{-6E_R}{k_B\theta_D}\left\{\frac{1}{4}+\left(\frac{T}{\theta_D}\right)^2\int_0^{\theta_D/T}\frac{xdx}{e^x-1}\right\}\right] \quad (3)$$

Where $E_R$ is the recoil kinetic energy, $k_B$ is Boltzmann constant.

In a thin absorber approximation *f* is proportional to a spectral area, *A*, so the latter can be used in a practical application of eq.(3) in order to determine $\theta_D$. For this purpose one calculates *ln (f/f$_o$)* as a function of temperature, where *f/f$_o$=A/A$_o$*, *A$_o$* being the spectral area at the lowest temperature measured (in our case 80K). The values of $\theta_D$ obtained by this method are displayed in Table 1. It is clear that they are significantly lower than those derived from the SOD approach. This feature is in line with a general findings e. g. for a metallic iron $\theta_D$(CS)=421(30) K while $\theta_D$(f)=358(18) K [12/13]. Important, however, is the fact that there is not difference, within the error limit, between the Debye temperature for the major and the minor phases viz. $\theta_D$ =207(13) K for the former and $\theta_D$=225(9) K for the latter as determined with the *f*-approach. This is rather unexpected result as the data reported in the literature give a clear evidence that the Debye temperature of the ferric iron (in the case of the Fe-gluconate usually associated with the minor phase) is significantly higher than the one of the ferrous compounds e. g. [11]. There are two plausible explanations for the present finding: either iron present in the minor phase is in the ferric–HS but its crystallographic structure is different (less rigid) than that of the major phase, or it is in the ferrous-LS and has the same or similar crystallographic structure as the major component i.e. the ferrous-HS. In these circumstances regarding the minor phase iron as ferric and suggesting that it is a product of oxidation of the ferrous iron e. g. [9] is very unlikely because this would mean that its structural conditions were the same as those of the ferrous iron. Consequently, the bounding of the ferric ions would be stronger than that of the ferrous ones, hence the Debye temperature of the former should be significantly higher. The present study gives evidence that this is not the case. On the other hand, recent multi techniques study of the Fe-gluconate gave an indication that some amorphous features were present [12]. If true, one could associate the minor phase with the amorphous structure. The Fe ions present in the latter could be ferric and less rigidly bounded as the Fe ions present in the major phase. The values of $\theta_D$ found in the present study would be in accord with such scenario.

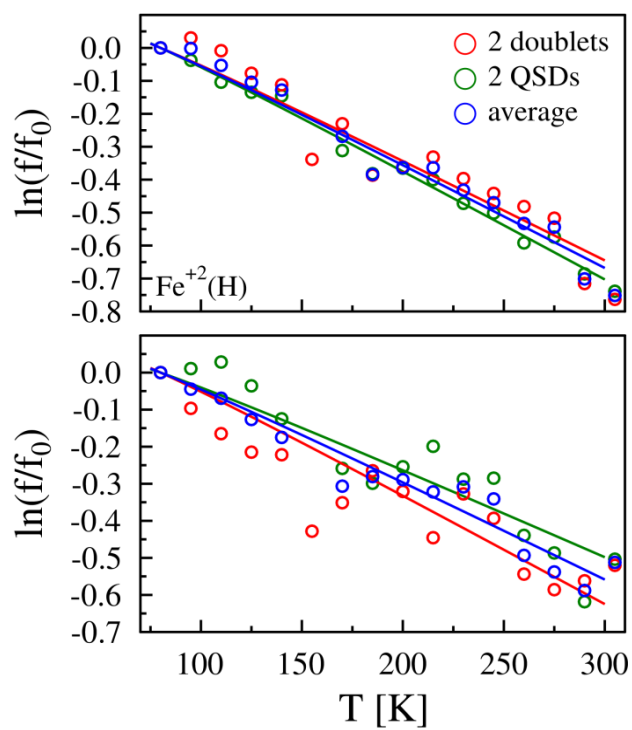

Fig. 3

Temperature dependence of ln(f/f₀), as found analyzing the spectra with the two fitting procedures as well as to the average arithmetic over the two sets of data. The solid lines stand for the best fit in terms of eq. (3).

### 3.3. Energetics of Fe atom vibrations

The knowledge of the SOD and the *f*-factor can be further used to determine average kinetic, $E_K$, and potential energy, $E_P$, of the vibrations, respectively. Concerning the former, it can be expressed as follows:

$$E_K = -mc^2 \frac{SOD}{E_\gamma} \qquad (4)$$

Where *m* stands for the mass of vibration atom and *c* for the velocity of light and $E_\gamma$ is the energy of gamma rays (14.4 keV in the present case).

The relationship between $E_P$ and *f* is given by the following expression:

$$E_P = -\frac{1}{2} D \left( \frac{\hbar c}{E_\gamma} \right)^2 \ln f \qquad (5)$$

Where *D* is a force constant.

The absolute values of $E_K$ can be readily calculated from eq. (4), while the absolute values of $E_P$ cannot as the value *D* is here unknown. Concerning the former, an increase of $E_K$ observed in the temperature interval of 80-305 K is equal to 22.4 meV. In Fig. 3 are presented relative changes of the average square velocity, <$v^2$>, as a function of the average square amplitude of vibrations, $<x^2>$. The former were calculated from *SOD* while the latter from *f*. It can be seen they are linearly correlated for the Fe atoms present in both phases. The increase of the two quantities is caused by the increase of temperature (from 80 to 305K). Some tiny anomaly can be seen in the mid part of the plot in Fig. $3a_1/a_2$ (it occurs at 180-190K). The present experiment, however, does not allow to conclude whether it is genuine or artefact.

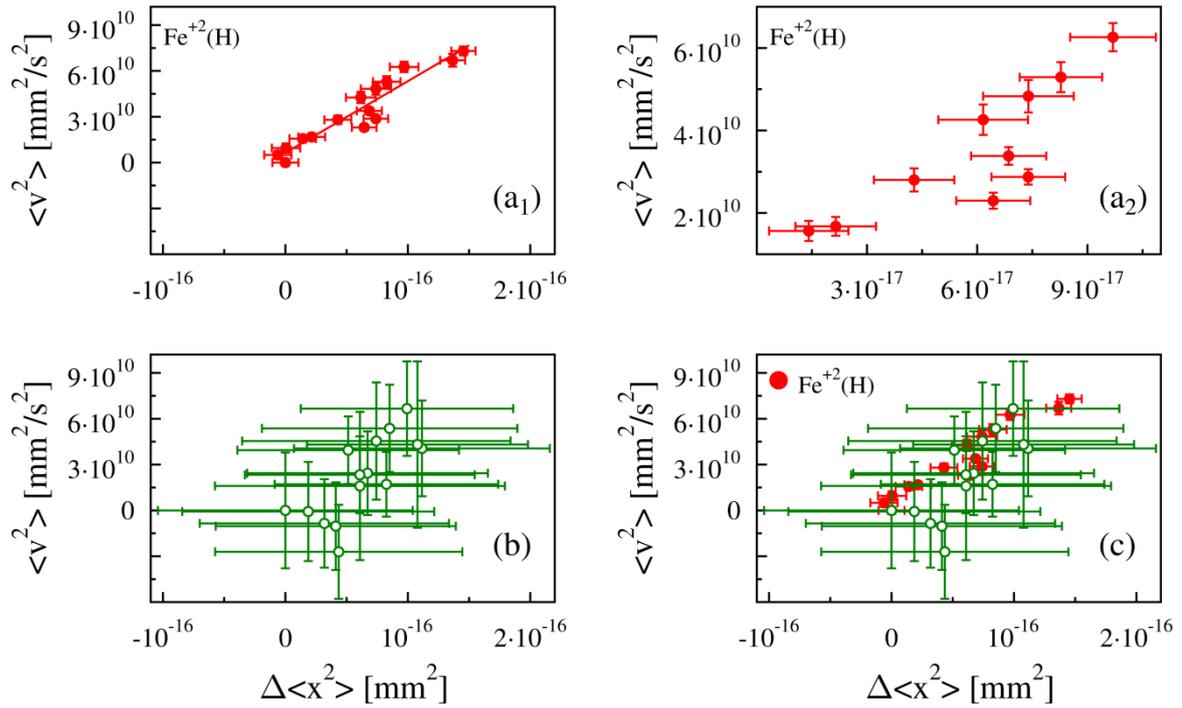

Fig. 4

The average square velocity, $<v^2>$, versus the average square amplitude, $<x^2>$, of Fe atom vibrations in the major (a1) and in the minor (b) phase of iron. A zoomed fragment of (a1) in the vicinity of a possible anomaly is shown in ($a_2$). Panel (c) illustrates a superposition of the data for both phases of iron.

The linear $<v^2>$-$<x^2>$ correlation shown in Fig. $3a_1$ can be further used to determine the bond force constant, *D*, for Fe atoms in gluconate. Assuming the vibrations are harmonic, $D = m\alpha$, where *m* is the mass of an $^{57}$Fe atom and $\alpha$ is the slope of the best linear fit to the $<v^2>$-$<x^2>$ data. In this way one obtains the value of *D*=44 N/m (243 cm$^{-1}$) which sounds reasonable. Its knowledge can be next used to determine a change of the potential energy, *ΔE$_P$ = 0.5kΔ<x$^2$>*, due to the increase of temperature from 80 to 305K. The value one arrived at is 20.6 meV which remains in a fair agreement with the corresponding increase of the kinetic energy. It should be added that the thermal energy related with the temperature increase by 225K amounts to 19.4 meV. It is evident that all three figures are consistent within ~10%.

**3.4. Temperature dependence of the quadrupole splitting**

Temperature dependence of the quadrupole splitting for the ferrous and the ferric ions was shown to be different [16]. For the former *QS* significantly decreases with *T* while for the latter the dependence is weak if any, so it can be used to make a distinction between the two forms of high-spin Fe ions . Figure 5 illustrates the present case. It is clear that the *QS (T)* determined from the major doublet (HS-ferrous ions) shows a significant *T*-dependence while *QS (T)* obtained from the minor doublet hardly depends on *T* as expected for the ferric state. The *QS (T)* dependence of the ferrous doublet could be satisfyingly described by the following equation:

$$QS(T) = QS(T_o)\left[1 - aT^{3/2}\right] \quad (6)$$

Noteworthy, the equation (6), whose origin is purely phenomenological and has not sound theoretical explanation, was successfully applied to describe the temperature *QS*-dependence of several classes of materials including crystalline, amorphous as well as quasi crystalline alloys e. g. [17,18].

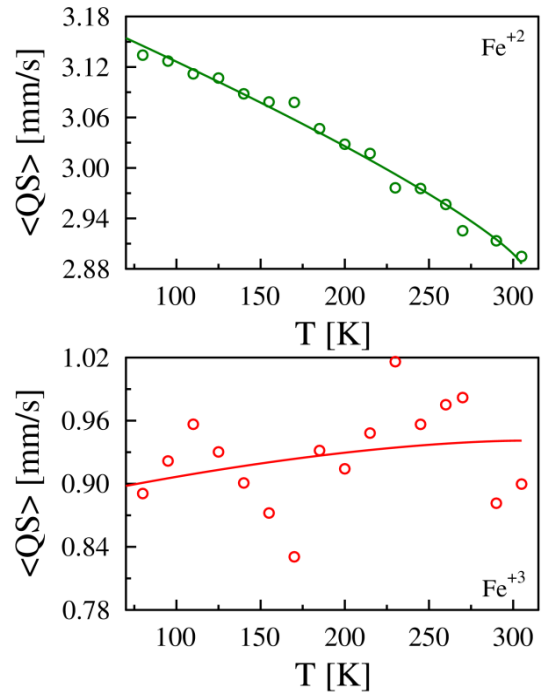

Fig. 5 Temperature dependence of the average quadrupole splitting, <QS>, for the major phase (upper panel) and that for the minor phase (lower panel). The solid lines represent the best fit to the data in terms of eq. (6).

## 4. Conclusions

The results obtained in the present study can be concluded as follows:

(1) Two phases of iron were detected in the Mössbauer spectra of Fe-gluconate: major (~85%) one unequivocally identified as containing high-spin ferrous ions, and minor (~15%) one whose identification was ambiguous as its spectral parameters (isomer shift and quadrupole splitting) overlap with those characteristic of a low-spin ferrous iron.

(2) Debye temperature was determined for both phases using temperature dependence of (a) the center shift and (b) relative spectral area. The former yielded the value of 445 (18) K for the high-spin ferrous phase and 357 (111) K for the minor phase, while the corresponding figures derived from the latter were 207 (13) K and 225 (48), respectively.

(3) Force constant for bonding of Fe atoms in the high-spin ferrous phase was calculated to be equal to 44 N/m or 243 cm$^{-1}$.

(4) The temperature dependence of the quadrupole splitting of the doublet associated with the high-spin ferrous phase follows the $T^{3/2}$ law, while the quadrupole splitting of the minor doublet hardly depends on $T$.

**Acknowledgements**

The Ministry of Science and Higher Education of the Polish Government is thank for financial support.